# Dynamics of transient metastable states in mixtures under coupled phase ordering and chemical demixing

Ezequiel R. Soulé[1] and Alejandro D. Rey[2]

1. *Institute of Materials Science and Technology (INTEMA), University of Mar del Plata and National Research Council (CONICET), J. B. Justo 4302, 7600 Mar del Plata, Argentina*

2. *Department of Chemical Engineering, McGill University, Montreal, Quebec H3A 2B2, Canada*

**Abstract.** We present theory and simulation of simultaneous chemical demixing and phase ordering in a polymer-liquid crystal mixture in conditions where isotropic-isotropic phase separation is metastable with respect to isotropic-nematic phase transition. In the case the mechanism is nucleation and growth, it is found that mesophase growth proceeds by a transient metastable phase that surround the ordered phase, and whose lifetime is a function of the ratio of diffusional to orientational mobilities. In the case of spinodal decomposition, different dynamic regimes are observed depending on the mobility ratio: metastable phase separation preceding phase ordering, phase ordering preceding phase separation, or simultaneous phase ordering and phase separation. Not only the overall dynamics but also the final structure of the material can be different for each kinetic regime.

# 1. Introduction

Rapid cooling of a liquid may result in a solid whose structure and composition is different when using slow cooling, even in the absence of vitrification, because transient metastable phases may emerge. For example in a mixture, liquid-liquid (LL) equilibrium can be buried below the liquid-solid (LS) transition, and with a fast and large undercooling, the metastable LL phase separation can precede crystallization [1]. The emergence of transient metastable phases and the evolution of phase transformation trough transient metastable states is known as the empirical Ostwald step rule [2]. Accordingly, crystallization consists of a sequence of chemical and structural changes rather than a single-step energy minimization step, known as kinetic crystallization pathway [1,2]. This is particularly relevant to crystallization of polymer solutions [1], where buried metastable states below the crystallization temperature are readily accessible through thermal quenches [3,4].

Traditionally, the main mechanism considered for the emergence of the transient metastable phase was its higher nucleation rate as compared with the stable one. In the last decade, Bechoeffer et al [5,6] described a new dynamic mechanism for the formation of metastable phases, based on the Landau-Ginzburg equation for a non-conserved order parameter (model A [7]). It was found [5,6] that an interface separating the stable high- and low- temperature phases can spontaneously split into two interfaces, one separating the high temperature phase and the metastable phase, and the other separating the metastable and the low temperature phase, independently of nucleation events. For a single order parameter, the necessary and sufficient condition for splitting is that the velocity of the second interface is higher that the velocity of the first one. In the case of more than one order parameter, the splitting can be hysteretic or need a finite magnitude perturbation, so the splitting may depend also on the initial conditions. A recent study in this area includes front splitting in the two order parameters of a thermotropic a smectic liquid crystal [8]. It was found, in qualitative agreement with experiments, that the structure of a growing smectic spherulite was modified by the process of front splitting.

The emergence of metastability by front splitting considered in the mentioned works is limited to non-conserved order parameters (NCOP). For mixtures, in addition to any NCOP, concentration (which is a conserved quantity) must be taken into account. Evans *et. al.*[9,10] analyzed the case of one conserved order parameter (COP), or model B [7]. They found that the single interface is stable with respect to splitting and a finite perturbation is needed in order to produce a metastable phase, and that this metastable phase can have a finite or infinite lifetime, depending on the degree of undercooling. In principle this type of model describes purely diffusional processes; but the minimum model for an order-disorder transition in mixtures must take into account one COP to describe diffusion and one NCOP to describe ordering (model C in [7]). As the dynamics of both order parameters is different, a more complex behaviour appears in these systems. Fischer and Dietrich [11] studied spinodal decomposition (SD) for model C and found different dynamic regimes depending on the relative values of diffusional and ordering mobilities. If ordering is much faster than diffusion, the system first evolves to a metastable homogeneous ordered state, and then phase-separates. When diffusion is comparable or faster than ordering, phase separation and phase ordering evolve simultaneously. Simulations [12,13] of polymer-liquid crystal mixtures exhibiting simultaneous phase ordering and chemical demixing have shown that liquid crystal orientational order interacts with chemical demixing producing morphologies sensitive to kinetic factors. The situation turns much more complex if metastable phases arise.

In a previous letter [14] we demonstrated that in simultaneous demixing and liquid crystal phase ordering, transient metastable states are formed during mesophase formation. Unlike the pure NCOP or COP case, each interface follows different kinetic laws (because the dynamic of conserved and non-conserved variables is different), and this fact lead to the result that the metastable phase has a finite lifetime. In this work, we extend the previous study of formation of multiple fronts, and analyze the different mechanisms of phase transition involving phase separation and phase ordering. Two characteristic and different cases are simulated: 1 – An interface between an isotropic and a nematic phase is taken as initial condition, representing growth of a nematic domain as in nucleation and growth (NG) mechanism, and the formation of multiple fronts is analyzed for different values of the two (diffusion and ordering) mobility ratios and different locations in the thermodynamic phase diagrams. 2 – A homogeneous system with small fluctuations, in the unstable region of the phase diagram, is taken as initial condition, and the dynamics of phase transition and structure formation through SD is analyzed again for different values of mobility ratio. The combined characterization of mobility effects with SD and NG modes gives a comprehensive picture of the kinetics of mixed order parameter mixtures.

The organization of this paper is as follows. Section 2.1 presents the model based on the Landau-deGennes/Cahn-Hilliard/Maier-Saupe/Flory-Huggins free energies, and briefly comments on the previously [14] used and validated numerical methods. Section 2.2 gives the sharp interface semi-analytical model used to determine interfacial velocities for diffusion and phase ordering, thus revealing their different time scalings. Section 3 presents and discusses the results. Section 3.1 presents growth of a pre-existing nematic phase, and Section 3.2 presents the spinodal decomposition model. Section 4 presents the conclusions.

## 2. Model

## 2.1 Landau-deGennes Phase Field Model

The free energy density of the system consists in homogeneous and gradient contributions. The Flory-Huggins theory is used for the mixing free energy, in combination with the Maier-Saupe theory for nematic order. The specific details of these theories can be found elsewhere [10-12,15]. The homogeneous free energy (per mole of cells) is:

$$\frac{f^{h}}{RT} = \frac{\phi}{r_c}\ln(\phi) + \frac{1-\phi}{r_p}\ln(1-\phi) + \chi\phi(1-\phi) + \frac{3}{4}\frac{\Gamma}{r_c}\phi^2 \mathbf{Q} : \mathbf{Q} - \frac{\phi}{r_c}\ln(Z) \quad (1)$$

where the partition function $Z$ is given by:

$$Z = \frac{1}{4\pi}\int_{0}^{2\pi}\int_{0}^{\pi}\exp\left[\frac{3}{2}\Gamma\phi\mathbf{Q}:\left(\boldsymbol{\pi}\boldsymbol{\pi} - \frac{\boldsymbol{\delta}}{3}\right)\sin^2\theta d\theta d\omega\right],$$

and $\phi$ is the liquid crystal volume fraction, the symmetric and traceless tensor $\mathbf{Q} = S\mathbf{nn} + P(\mathbf{ll} - \mathbf{mm})$ is the quadrupolar order parameter [12], (which plays the role of a "phase field") where $S$ and $P$ are the scalar uniaxial and biaxial parameters, $\mathbf{n}$, $\mathbf{m}$ and $\mathbf{l}$ are the eigenvectors of $\mathbf{Q}$, $r_c$ and $r_p$ are the ratios of molar volume of liquid crystal and polymer with respect to the cell volume, $\chi$ is the mixing interaction parameter, $\Gamma$ is a Maier-Saupe interaction parameter (both are functions of $1/T$), $\boldsymbol{\pi}$ is a unit vector and $\boldsymbol{\delta}$ is the identity matrix, $R$ is the gas constant and $T$ is the temperature. $Z$ was accurately approximated by a polynomial expression in terms of the invariants of $\mathbf{Q}$ as $Z = a(\mathbf{Q}:\mathbf{Q}) + b\left[(\mathbf{Q}\cdot\mathbf{Q}^T):\mathbf{Q}\right] + c(\mathbf{Q}:\mathbf{Q})^2 + d(\mathbf{Q}:\mathbf{Q})\left[(\mathbf{Q}\cdot\mathbf{Q}^T):\mathbf{Q}\right]$. The coefficients $a$, $b$, $c$ and $d$ were obtained from a least-squares fitting of the numerical solution of the integral, as described in ref 16. The gradient free energy is given by gradients in concentration and order [11-13]:

$$f^g = l_\phi\left(\nabla\phi\right)^2 + l_{Q1}\nabla\mathbf{Q}:\nabla\mathbf{Q} + l_{Q2}\left(\nabla\cdot\mathbf{Q}\right)\cdot\left(\nabla\cdot\mathbf{Q}\right) + l_{Q\phi}\left(\nabla\cdot\mathbf{Q}\right)\cdot\nabla\phi \tag{2}$$

The time-dependent Ginzburg-Landau formulation and the Cahn-Hilliard equation are used to simulate the time evolution of $\mathbf{Q}$ and $\phi$ [13]:

$$\frac{\partial\mathbf{Q}}{\partial t} = M_Q\left(-\frac{\partial f}{\partial\mathbf{Q}} + \nabla\cdot\frac{\partial f}{\partial\nabla\mathbf{Q}}\right) \tag{3}$$

$$\frac{\partial\phi}{\partial t} = M_\phi\nabla^2\left(\frac{\partial f}{\partial\phi} - \nabla\cdot\frac{\partial f}{\partial\nabla\phi}\right) \tag{4}$$

The numerical methods used to compute phase diagrams are given in [12,13]. Comsol Multiphysics was used to solve eqn.(3,4), with quadratic Lagrange basis functions; standard numerical techniques were used to ensure convergence and stability.

## 2.2 Sharp-Interface Semi-analytical model

The overall dynamics of the system is given by the coupled equations 3 and 4, but in the case that one variable is much slower than the other, the overall kinetics will be controlled by this slower variable. The two limiting cases of kinetics controlled by diffusion and by ordering can be analyzed with simplified semi-analytical models, as follows. These simplified models can be used to analyze the simulation results and determine if one mechanism dominates the dynamics or if there is a transition from one mechanism to the other.

*(A) Diffusional kinetics:* diffusion can be analyzed in terms of Fick´s law, which is a sharp-interface equivalent to Cahn-Hilliard model if the diffusivity is $D = M_\phi \partial^2 f / \partial \phi^2$ and the gradient terms (which are expected to be important only in the interface) are neglected in the bulk phases. The assumption of constant mobility implies a non-constant diffusivity, but in order to find an analytical solution it will be assumed to be constant, evaluating the second derivative of the free energy at an average concentration between the interface and the bulk. The boundary and initial conditions are

$$\phi\big|_{x=X_{INT}} = \phi^i_{INT}$$

$$\phi\big|_{x=\infty} = \phi^i_{bulk} \tag{5}$$

$$\phi\big|_{t=0} = \phi^i_{bulk}$$

$$(\phi^n_{INT} - \phi^i_{INT})v = D\partial\phi/\partial x$$

where $X_{INT}$ is the position of the interface, $\phi^i_{bulk}$ is the concentration in the isotropic bulk phase, $\phi^i_{INT}$ and $\phi^n_{INT}$ are the concentration in the nematic and isotropic sides of the interface (equilibrium concentrations), and $v$ is the velocity of the interface. The concentration profile can be found using a similarity transformation [17], and is given by:

$$\phi = \frac{\phi^i_{bulk} - \phi^i_{INT}}{1 - erf(\sigma)} \left[ erf(\eta) - erf(\sigma) \right] + \phi^i_{INT} \tag{6}$$

where $erf(x)$ is the error function and the factor $\sigma$ satisfies:

$$2\left(\phi^n_{INT} - \phi^i_{INT}\right)\sigma = \frac{\phi^i_{bulk} - \phi^i_{INT}}{1 - erf(\sigma)} \exp\left(-\sigma^2\right) \tag{7}$$

The well-known velocity of the interface $v = \sigma D^{1/2} t^{-1/2}$, and it slows down with elapsed time as $1/\sqrt{t}$.

*(B) Phase Ordering Kinetics*: an expression for the velocity can be found following the procedure of ref. [18]. If the process is kinetically controlled by ordering the front velocity is $v = \frac{1}{\beta} \mathbf{L}$, where $\beta = \frac{1}{M_Q} \int\limits_{\Delta N}^{\Delta I} \partial \mathbf{Q} / \partial x : \partial \mathbf{Q} / \partial x \, dx$ is the interfacial viscosity, $L = \int\limits_{\Delta N}^{\Delta I} \partial f / \partial \mathbf{Q} : \partial \mathbf{Q} / \partial x \, dx$ is the bulk stress load, and $\Delta I$ and $\Delta N$ are the positions of isotropic and nematic boundaries of the interface. It has been observed in the simulations that the change in $\mathbf{Q}$ across the interface is more abrupt than the drop in $\phi$, so the integrand in the expression for $L$ is non-zero in a region where $\phi$ varies slightly and it's close to the nematic phase value. We can assume that all the change in $\mathbf{Q}$ is located in a region where $\phi$ is constant (this will give the maximum driving force for ordering, which is reasonable considering that the kinetics is controlled by ordering), so we can calculate the bulk stress load as the change in free energy: $L = f(\mathbf{Q}^n, \phi_{INT}^n) - f(0, \phi_{INT}^n)$. The interfacial viscosity $\beta$ defined above can be found by numerical integration of the order parameter profiles found in the simulations.

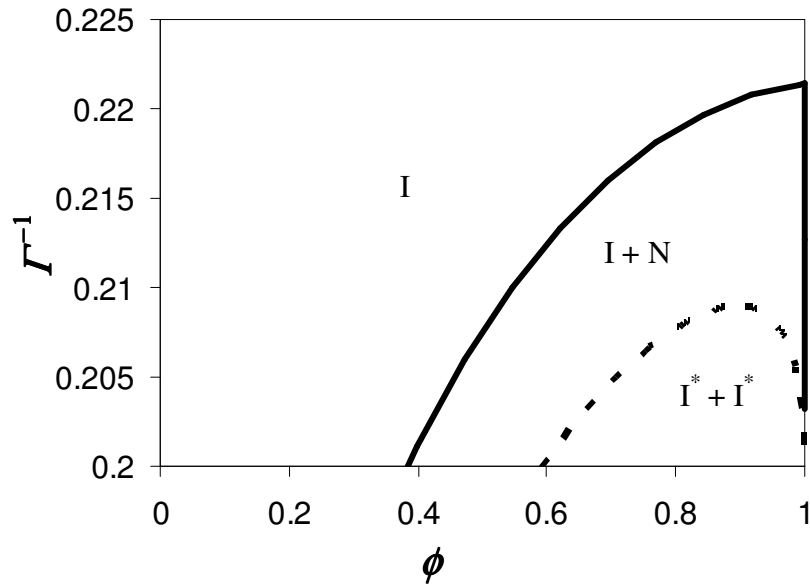

**Figure 1.** Computed phase diagram based on eqn. 1; for: $r_c$=2.74, $r_p$ =200, $\chi$=-.645+.18/$\Gamma$. $\Gamma$ is the inverse of the nematic interaction parameter and $\Gamma^{-1}$ is a dimensionless temperature, and $\phi$ is the volumetric fraction of liquid crystal. I: isotropic, N: nematic, I*: isotropic metastable.

## 3. Results and Discussion

Figure 1 shows the thermodynamic phase diagram used in this study computed using eqn. (1). The construction of this phase diagram follows standard procedures: the equilibrium condition at each temperature is given by the equality of chemical potentials of each component and the minimization of the free energy with respect to the order parameter, in each phase. A more detailed analysis of phase diagrams can be found in [15]. Two types of phases exist: isotropic ($\mathbf{Q} = 0$) and nematic ($\mathbf{Q} > 0$). An I + I coexistence region exists within the N+I coexistence region, and thus it is a metastable equilibrium (this denoted by I*+I* in Fig.1). This means, the free energy is minimal when I and N phases coexist, but it has a local minimum for I + I coexistence. The issue to be established is how the metastable gap influences the phase ordering-demixing process.

### 3.1 Growth of a pre-existent nematic phase

Partial results for the growth of a nematic phase into the isotropic phase have been introduced in our previous letter [14]. The representative initial condition is an interface separating an isotropic phase with a given $\phi_{bulk}^i$ from a nematic phase with $\phi$ and $\mathbf{Q}$ given by the equilibrium conditions at the temperature under consideration. This can represent, for example, at system with concentration $\phi_{bulk}^i$ that is quenched to a temperature in the coexistence region in the phase diagram, where a domain of nematic phase has been nucleated. As this is a 1D system and there is no change on the orientation in space, the analysis of nematic order can be done in terms of the scalar order parameter, $S$. The dynamic evolution of the system represents the growth of the nematic phase. Different values of $\phi^i_{bulk}$ and diffusional-to-phase ordering mobility ratio $M_R = M_\phi/l^2 M_{\mathbf{Q}}$ were used, where $l$ is the characteristic length, defined as $l = (l_\phi/RT)^{1/2}$. The spatial position is expressed in units of this characteristic length and the time is expressed in units of $\tau = (M_Q RT)^{-1}$. The gradient parameter ratios $l_{Q1}/l_\phi = l_{Q2}/l_\phi = 0.1$ and $l_{\phi Q}/l_\phi = 0.5$ were used in all simulations (see equation 2).

For high values of $M_R$ (diffusion is faster than ordering), the interface was observed to spontaneously split in two, one separating the two metastable isotropic phases, and the second one separating an isotropic phase and the equilibrium nematic phase, as shown in fig. 2a. This is in agreement with [5,6,8,9]: the whole interface is considered as being composed by two interfaces: nematic – isotropic 1 and isotropic 1 – isotropic 2. As ordering is slow, the dynamics of the first interface will be controlled by ordering and it will be slow, whereas the second interface has a pure diffusional dynamics and it moves faster, so the interface splits in two. For NCOP [5,6] both interfaces have constant velocity (ordering dynamics), and the distance between the two interfaces remains constant or increases linearly with time. For COP [8,9], this distance can increase indefinitely or the interfaces can merge again, and this behaviour depends only on the degree of undercooling. In the present mixed order parameters case, the I-N interface has a constant velocity but the I-I interface slows down with time, so at short times the interfaces separate but after some time they merge again, as shown in fig. 2.b.

Thus, under phase ordering control the metastable phase has a finite lifetime, as in the Ostwald step rule [1,2].

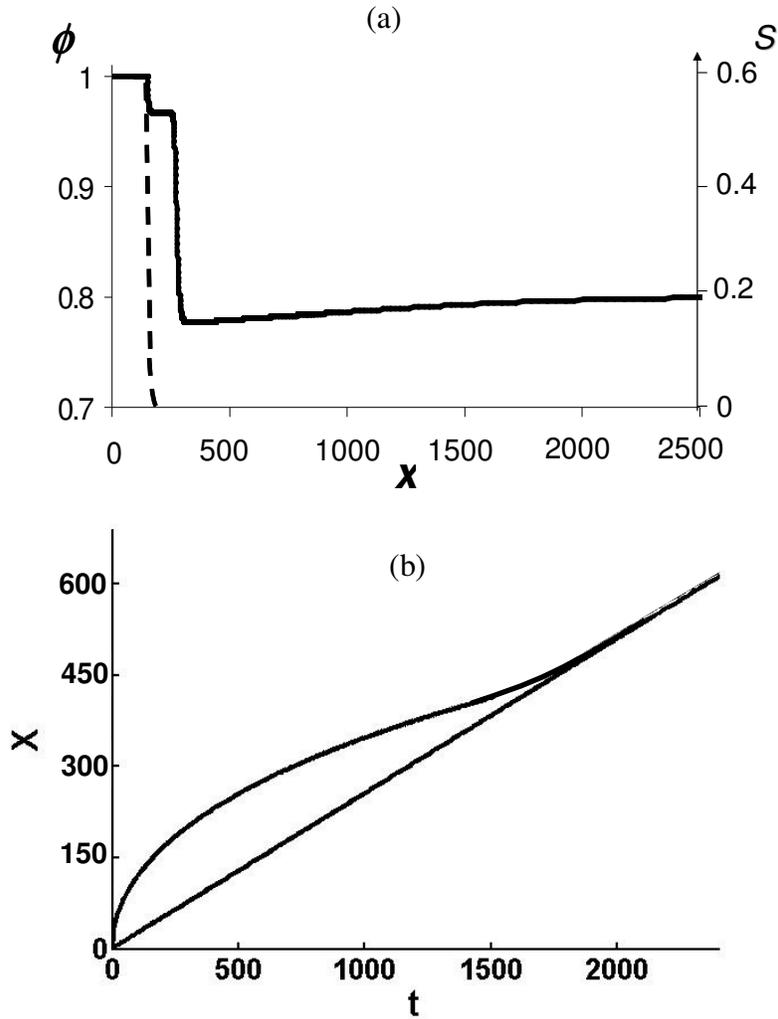

**Figure 2.** Interface splitting observed with $M_R = 2.4 \cdot 10^6$. (a) Concentration $\phi$ (full line, left axis) and scalar order parameter $S$ (dashed line, right axis) profiles at $t = 650$. $X$ represents the dimensionless position (b) Dimensionless position of the two interfaces shown in (a) (dashed line: N-I interface; full line: I-I interface), as a function of dimensionless time.

As this is a system of two different order parameters, the splitting-merging behaviour depends on the mobility ratio (as opposed to the pure NCOP or COP case were it is controlled only by the degree of undercooling). When the value of $M_R$ decreases the lifetime of the metastable phase decreases. A theoretical plot of the metastable phase lifetime versus $M_R$ was constructed from the simplified sharp-interface model (see section 2.2) and compared with the simulations. The velocity of the nematic/isotropic interface can be calculated assuming that its kinetics is controlled by ordering, with $L/RT$=0.0162 and $\beta/M_Q l = 0.0663$, giving $v_{N-I} = 0.23$. The velocity of the isotropic/isotropic interface can be approximated solving Fick's law, with $\phi_{bulk} = 0.8$ and $\phi_{INT} = 0.776$, giving $v_{I-I} = 3.78 M_R^{1/2} t^{-1/2}$. The time at which both interfaces intersect each other is given by $t = 4.9 \cdot 10^{-3} M_R$. This is plotted in fig. 3, together with the merging times observed in the simulations.

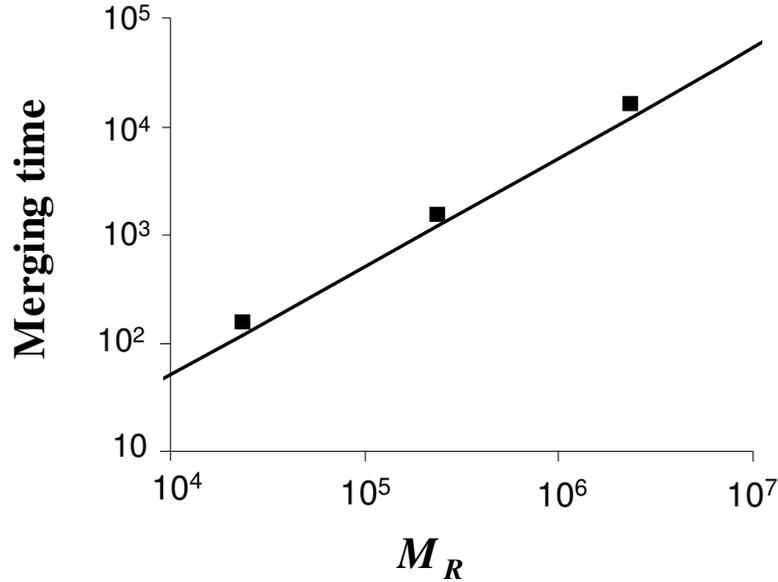

**Figure 3**. Merging time as a function of mobility ratio, from numerical simulations (squares) and from eqn. 4 (full line). As $M_R$ increases the lifetime of metastable state increases.

For sufficiently small mobility ratio the merging time will be so small that the maximum separation between the interfaces will become smaller that the interface width, so there will be an "incomplete" splitting. The separation between both interfaces, according to the simplified model, is $\Delta x = 8.11 M_R^{1/2} t^{1/2} - .23t$, and the maximum separation is $\Delta x = 3.25 \cdot 10^{-4} M_R$. For a mobility ratio of $2.4 \cdot 10^4$, the maximum separation is 7.8, and the interface thickness (observed in simulations) is about 20, so "incomplete" splitting is expected. This was observed in simulations, where splitting was not seen, but a shoulder appeared in the interface at short times and then disappeared. This can be seen in fig 4, which is a plot of the concentration profiles across the interface at different times. The merging time in this case was taken as the time at which the shoulder disappears completely.

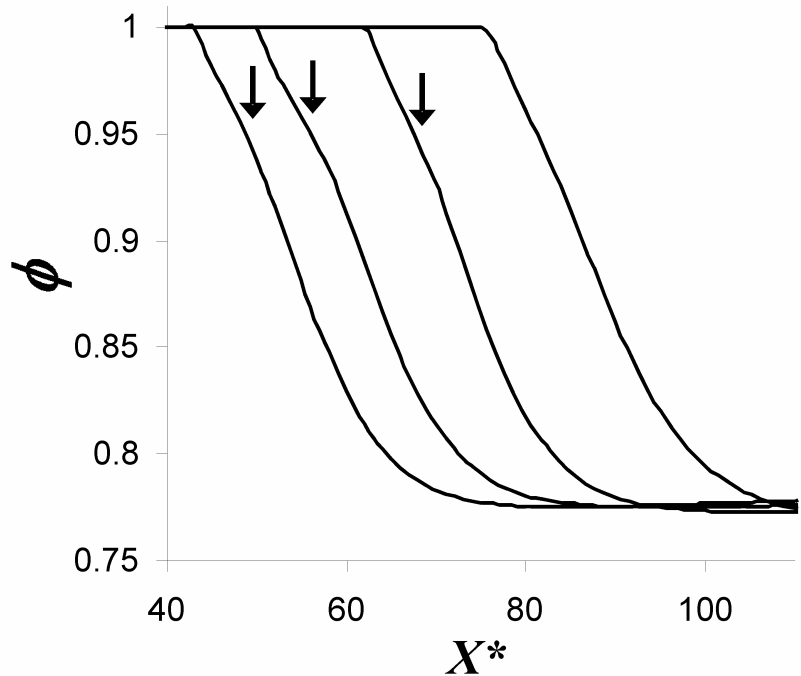

**Figure 4.** Concentration profiles across the interface for $M_R$=2.4·$10^4$ for dimensionless times (curves from left to right): 20, 60, 120, 180. $\phi$ is the volumetric fraction of liquid

crystal and $X$ the dimensionless position. The arrows indicate the location of the shoulder, as discussed in the text.

Finally, we studied how the interface splitting-merging process depends on the location of the mixture in the phase diagram. Different concentrations were analyzed for the same temperature and the same mobility ratio, and the results of merging time are shown in figure 5. It was observed that as the mixture composition approaches the binodal I-I composition, (as the concentration of the interface approaches the concentration of the metastable phase) the merging time goes to zero. This can be seen analytically in eqn. 7, as $\phi_{bulk}^{i}$ - $\phi_{INT}^{i}$ goes to zero, $\sigma$ goes to zero and the velocity of the interface goes to zero, so no interface splitting takes place. On the other hand, as the initial concentration is deep into the binodal region, the larger is the velocity of the diffusional interface and the larger is the lifetime of the metastable phase. As the spinodal composition is approached, and the second derivative of the free energy changes significantly with the composition, the assumption of Fickian diffusion with constant diffusivity no longer holds and the sharp-interface model results deviate from simulations. Actually, the concentration profile in the diffusional depletion layer becomes almost a straight line, as opposed to the common profile with smoothly changing curvature observed in the other cases. For concentrations inside the spinodal line, spinodal decomposition becomes possible competing with nucleation and growth, so this analysis was not extended to this condition. The case of spinodal decomposition is analyzed in the next section.

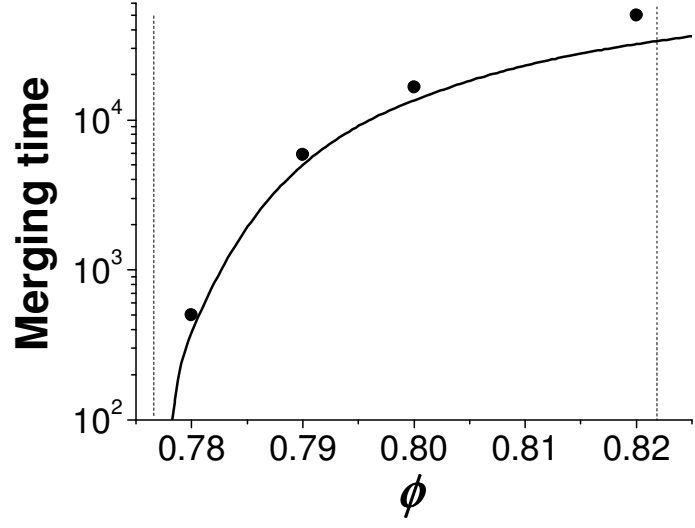

**Figure 5**. Merging time as a function of concentration from simulations (squares) and from eqn. 4 (line). The mobility ratio used was $2 \cdot 10^6$. The limiting cases of binodal composition and spinodal composition are shown with dotted lines.

Having analysed the splitting-merging mechanism, next we establish the post-merging kinetics. The formed single interface is found to present a dynamics that is neither purely ordering nor purely diffusional. Figure 6 shows the results from the simulation with $M_R = 2.4 \cdot 10^4$ (small merging time – incomplete splitting). We also plot the limiting velocities calculated with the simplified models for ordering- and diffusion-control. It can be seen that, at short times the kinetics is closer to be ordering-controlled, but as time goes on, there is a transition to diffusional control. This is because ordering kinetics is independent of time, while diffusion slows down as the concentration profiles develop, so eventually, at some time the kinetics must become diffusion-controlled.

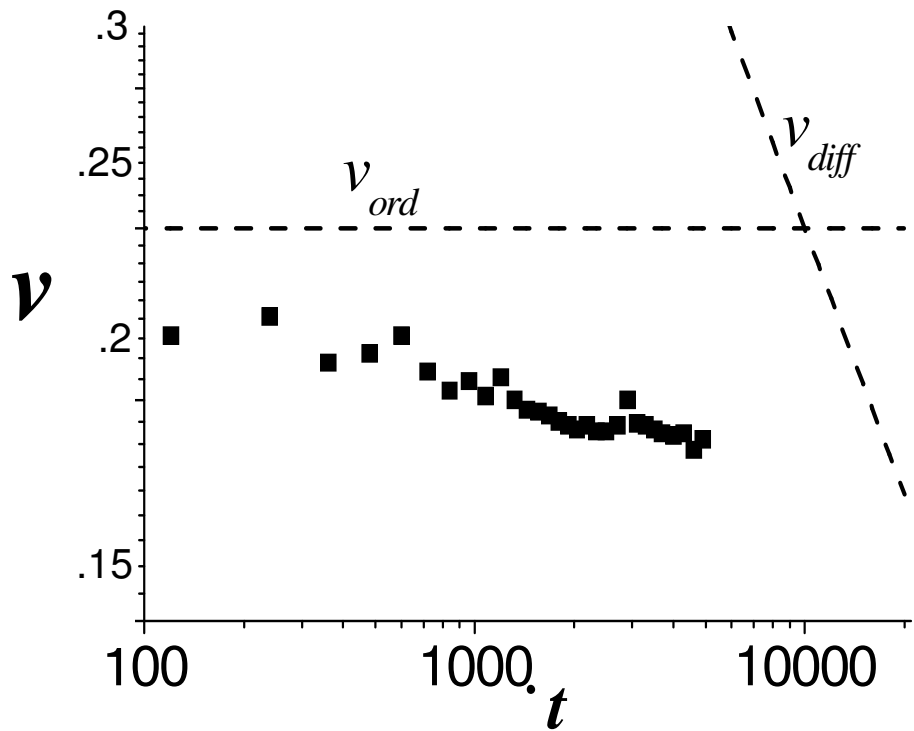

**Figure 6.** Dimensionless interfacial velocity as a function of dimensionless time; squares are from simulations, and the dashed lines are calculated with the simplified diffusion and phase ordering models given in Section 2.2.

As shown in fig. 2a, the concentration drop at the nematic-isotropic interface when the metastable phase forms is small, in particular it is smaller than when a unique interface is present. This implies a different value of interfacial tension. As texturing and defect dynamics depends on interfacial tension, the formation of a metastable phase is expected to have an impact in texture and defect formation for bi- or tri-dimensional geometries (as observed in ref. 12), and thus it will affect the optical properties of the material. Furthermore, this phenomenon is not restricted to polymer-liquid crystal mixtures; it could be present in any system with coupled conserved and non-conserved order parameters (e.g.: metallic alloys, dispersions of particles forming colloidal systems, etc). Another implication that arises from our analysis is that, even when the overall kinetics of the process will be diffusion-controlled at long times, the dynamic behaviour is more complex at short or intermediate times, and this can affect the final morphologies as well as the overall time scale of the process.

## 3.2 Spinodal Decomposition

Some time ago, Fischer and Dietrich studied spinodal decomposition in a model with mass diffusion and a non-conserved order parameter, similar to the one analyzed in this work [11]. The main difference is that they didn't consider the possibility of a metastable phase separation. They found two different regimes, depending on the mobility ratio: for low mobility ratio (fast ordering), phase ordering precedes phase separation, so the system becomes ordered while remaining homogeneous in a first stage (with a metastable value of order parameter that minimizes the free energy for the concentration of the mixture), and in a second stage the system phase separates, and the order parameter relaxes to the equilibrium values in each phase as the phases are formed. For high mobility ratio (fast diffusion), phase separation and phase ordering are simultaneous, and the spatial and temporal changes in order parameter and concentration are fully coupled during the whole process. In the present system, as will be described next, there is a new regime, corresponding to phase separation leading to the metastable state in a first stage, and phase ordering-phase separation leading to the equilibrium state in a second stage.

Spinodal decomposition in one dimension was analyzed for a mixture with $\phi = 0.95$, $S=0$ and $\Gamma^{-1}=0.187$, and for three very different values of relative mobility, corresponding to fast ordering, fast diffusion and an intermediate case. In these conditions the system is unstable to infinitesimal variations of both composition and order parameter. A random Gaussian noise of zero mean and $10^{-3}$ standard deviation was added to the initial condition in order to initiate the spinodal decomposition (the absolute value of the noise was taken for the nematic order parameter, to avoid negative values). Periodic boundary conditions were used. Several simulations with different initial noises were performed for each value of relative mobility.

Figure 7 show representative concentration and order parameter profiles at different times for each value of relative mobility (corresponding to one quarter of the total size of the simulation domain), and their corresponding structure factors (averaged on all the simulations performed for each case), are presented in the appendix (supplementary material). It can be seen that the evolution of phase transition is very different for each case, as will be analyzed next.

When the mobility ratio is small (figures 7a), phase ordering precedes phase separation, so the mixture first evolves to a metastable homogeneous nematic state (t<350). At t=390, fluctuations of concentration located at the walls of the nematic domains start to grow significantly. As the equilibrium value of the order parameter depends on the concentration of the mixture, the evolution of the order parameter couples to the concentration and the formed metastable nematic domains break down into subdomains of the stable phases (t=50000). As the mixture keeps evolving towards equilibrium the domains coarsen (t=400000) in order to decrease surface energy.

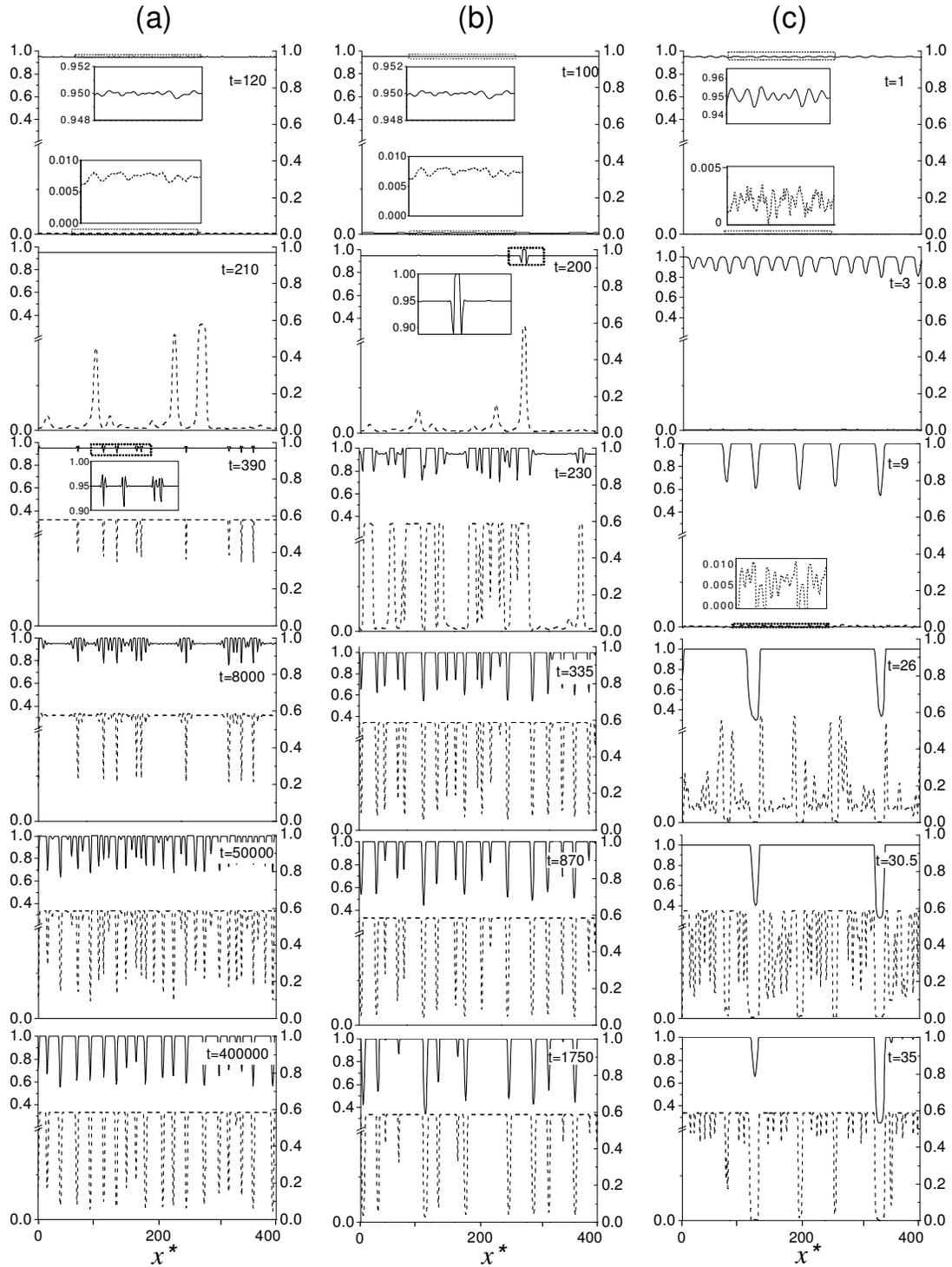

**Figure 7**. Evolution of concentration (full line, left hand axis), and scalar order parameter (dashed line, right hand axis) profiles for $M_R = 10^{-3}$ (a), 1 (b), and $10^3$ (c). The inlets figures correspond to zooms in the areas indicated by dotted rectangles.

For an intermediate value of mobility ratio (figs 7b), fluctuations in concentration and order parameter evolve simultaneously growing first and coarsening later.

For the case of high mobility ratio (figures 7c), again the process is step-wise. First, the system phase-separates in the metastable isotropic phases, with no significant evolution of the order parameter, through spinodal decomposition (t<9). At some point, the nematic order parameter start to increase within the high concentration regions (t=9 and 26), and ordered sub-domains are formed (t=30.5). At later times, the nematic sub-domains start coarsening (t=35).

It should be pointed out that, in these 1-D simulations, coarsening of nematic domains can be complete as there are no defects or orientation conflicts. But in a 2D or 3D system, different kind of defects can be formed when two domains become into contact. The domain structures and the different defects can then have long lifetimes or be metastable. Taking this into account, it is interesting to observe that, in the case of high mobility ratio, the structure of nematic sub-domains within the larger domains formed during phase separation can be stabilized for long times and thus can become the final structure of the material, unlike the case of intermediate mobility ratio, where a subdomain structure is not formed, or the case of small mobility ratio where there is a transient sub-domain structure but it rapidly breaks down completely when phase separation starts.

The characterization of the dynamics and of the multiple stages found in figure 7, is efficiently performed by considering the time evolution of the maximum structure factor and wavenumber corresponding to the maximum, $k_m$, shown, respectively, in the upper and lower panels in Figure 8. We have identified four different stages in each of the three cases shown in fig.7. For $M_R = 10^{-3}$ (fig. 8a), stage 1 (t<220) corresponds to an early stage of ordering process ($k_m$=0, exponential growth of structure factor), with no significant evolution of phase separation. In stage II, 220<t<400, phase separation starts to evolve (early stage, with exponential growth and constant $k_m$). A maximum appears in the nematic structure factor (domain structure), and it moves to higher values of $k$ with time (as phase separation evolves, new smaller nematic domains are formed). In the third stage, 400<t<9000, the morphology of nematic domains keeps refining, ($k_m$ keeps increasing), and the fluctuations in concentration keep increasing and coarsening, (so $k_m$ decreases). The structure factor keeps increasing, but with a power law rather than exponentially (note the change in the time scale from linear to logarithmic in fig. 8). Finally, the last stage for t>10000 represent the late stage of the process, where now both ordering and concentration are fully coupled, and follow a coarsening dynamics characterized by power laws. The observed exponent for $k_m$ is $n \approx 0.2$ which is lower than the expected 1/3 for this process, indicating that the asymptotic long time regime has not been reached yet.

The results for $M_R = 1$ are shown in fig. 8b. In the early stage t<220, both structure factors increase exponentially. For 220<t<300, the nematic structure factor starts to develop a peak and the maximum value starts to decrease. Localized zones where the phase transition has taken place can be distinguished from zones where the process has still not advanced (see figure 7b), and $k_m$ for the concentration decreases but with no clear exponential or power law. For 300<t<420, all the domains are formed, and $k_m$ is coincident for both structure parameters (see appendix for details). In the final stage, for t>420, the domains coarsen and $k_m$ decreases with a power law ($n \approx 0.19$), and the structure factors increase with approximately a power law ($n = 0.5$-$0.55$). Again, the expected value of $n=1/3$ for $k_m$ was not reached for the times analyzed.

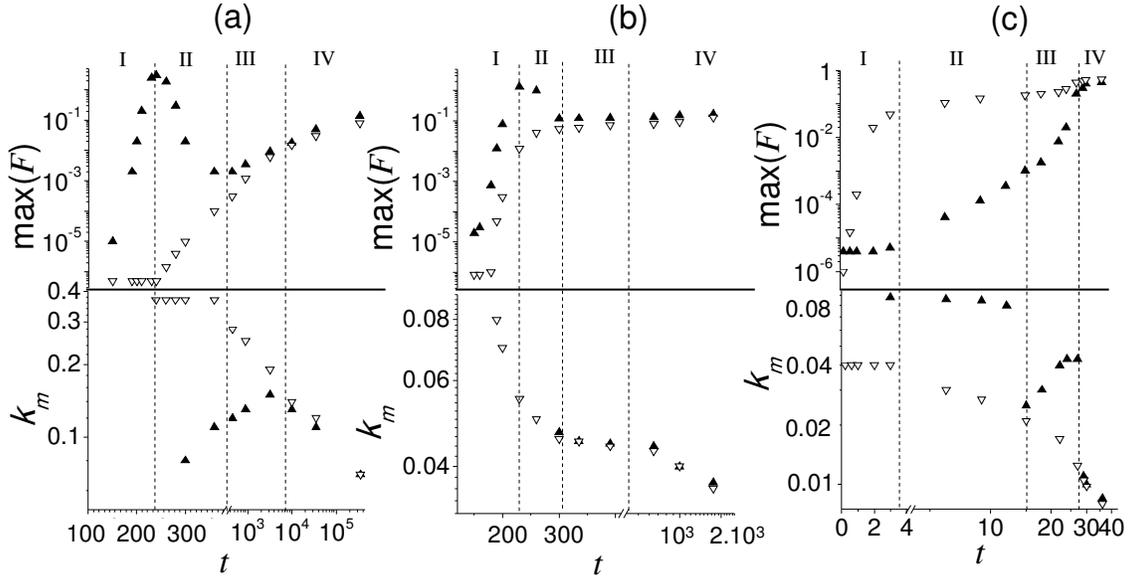

**Figure 8**. Maximum value the structure factor (upper), and wavenumber at which the maximum is located (lower), as a function of time for MR = $10^{-3}$ (a), 1 (b), and $10^3$ (c). Full symbols correspond to nematic structure factor, void symbols to concentration structure factor. Note that the time scale changes from linear to logarithmic.

Finally, fig. 8c shows the dynamics for $M_R = 10^3$. The first stage for t<2.5 corresponds to the early stage of spinodal decomposition, with the structure factor increasing exponentially with a fixed $k_m$. The second stage, 2.5<t<12, corresponds to an intermediate-late stage of phase separation and an early stage of phase ordering. The nematic structure factor grows exponentially, while the concentration structure factor evolves with power laws; $k_m$ decreasing with an exponent close to $n = 1/3$ as expected in a coarsening process. The nematic structure factor has multiple peaks (see appendix), wiht the maximum (dominant lengthscale) corresponding to the fluctuations within the high concentration domains. The third stage corresponds to an intermediate regime of phase ordering, coupled to concentration. There is a change in the dominant length scale, at the beginning of this stage $k_m$ is coincident with the concentration lenghtscale, but it increases as the subdomain structure develops (see figure 7c and appendix). The concentration evolves from the metastable equilibrium to the stable equilibrium resulting in an increase in the apparent growth exponents (see the change in slopes in figure 8c). Finally the last stage is a coarsening process. $k_m$ for both structure factors is coincident, although the sub-domain structure still persists. The apparent exponent for $k_m$ is still significantly higher that 1/3, meaning that the asymptotic value has not been reached yet, but we expect that, as the sub-domain structure disappears and the order parameter fully couples to concentration, the coarsening growth law with $n = 1/3$ should be reached.

In principle, we would expect that the main features of the observed behaviours (different regimes and the dependence of the dynamics and the structures formed on the mobility ratio), remain in higher dimensions, as they depend on the competition between the dynamics of ordering and diffusion and the formation of metastable phases. We anticipate that the situation can be much more complex due to the presence of curvature, defects, textures and different types of morphologies (co-continuous vs droplet).

## 4. Conclusion

The dynamics of an isotropic-nematic phase transition in a binary system exhibiting simultaneous chemical demixing and liquid crystal phase ordering was simulated, in conditions where a metastable isotropic-isotropic phase equilibrium exists. The effect of this metastable phase separation in the overall dynamics was analyzed for two cases: pre-existing nematic-isotropic interface (this case would be analogous to the mechanism of nucleation and growth), and spinodal decomposition.

In the first case, the transformation evolves via multiple stable-metastable interfaces that arise due to the distinct kinetics of phase ordering and diffusional fronts. For high mobility ratios the interface spontaneously splits in two, one separating the two metastable isotropic phases, controlled by diffusion, and the other one separating one isotropic and one nematic phase, controlled by ordering, in accordance with the mechanism found by [5,6] for non-conserved order parameters. But in the present case, as the dynamics of both interfaces are different, the interfaces merge again after some time, as observed for conserved order parameters [9,10], so the metastable phase has a finite lifetime following Ostwald step rule [1,2]. After merging, the interface has a mixed dynamical behavior, nor purely ordering-controlled nor diffusion-controlled. At the beginning the velocity is close to the ordering-controlled velocity, but as diffusion slows down with time, the system transitions to diffusional control and the velocity decreases with time.

In the case of spinodal decomposition, several regimes were observed depending on the mobility ratio for ordering and diffusion. When ordering is faster than diffusion, the system first evolves to a metastable ordered state with uniform composition, and then phase-separates into the two stable phases, as observed previously by Fischer and Dietrich [9]. For intermediate mobility ratios, phase ordering and phase separation evolve simultaneously and the profiles of order parameter and concentration are fully coupled throughout the whole process. When diffusion is faster than order, the system first phase separate into the metastable isotropic-isotropic equilibrium, and then phase ordering and phase separation takes place within the high-concentration domains. The dynamics of the system was analyzed in terms of the evolution of the maximum value of the structure factor and the wavenumber at which it is located. It was found that, in general, the different stages of the process can be described with simple exponential or power law growth laws, but the sequence and combinations of different laws for the two variables in the different stages depend on the value of mobility ratio, giving rise to a very rich dynamic behaviour.

**Acknowledgement**

ADR acknowledges partial support from the Natural Science an Engineering Research Council of Canada (NSERC) and the Petroleum Research Fund. ERS acknowledges a scholarship provided by the National Research Council of Argentina (CONICET).

## Appendix

The purpose of this Appendix is to present the evolution concentration and nematic structure factors, for the cases of spinodal decomposition analyzed in section 3. These figures, in conjunction with figures 7 and 8 provide a comprehensive description of the process of spinodal decomposition.

Three different structure factors can be defined (see ref [11]) to characterize fluctuations in concentration, order parameter, or both:

$$F_\phi(k) = \left| \sum_j [\phi(x) - \phi_0(x)] \exp(2\pi i k x_j) \right|^2$$

$$F_S(k) = \left| \sum_j S(x) \exp(2\pi i k x_j) \right|^2$$

$$F_{S\phi}(k) = \left| \sum_j \left[ \phi(x) - \phi_0(x) \right] \exp\left(2\pi ik x_j\right) \right| \left| \sum_j S(x) \exp\left(2\pi ik x_j\right) \right|$$

Figure A1 shows the structure factors for $M_R = 10^{-3}$, corresponding to the profiles shown in figure 7a. As discussed in section 3, at short times phase ordering evolves at constant concentration (fig A1.a), but for t>200 phase separation starts to evolve and the structure factor increases, showing a maximum, that moves to smaller values of $k$ as the process evolves. As the domain structure becomes noticeable the nematic structure factor starts to develop a maximum too (Fig A1.b), that after some time becomes coupled with the maximum of the concentration structure factor.

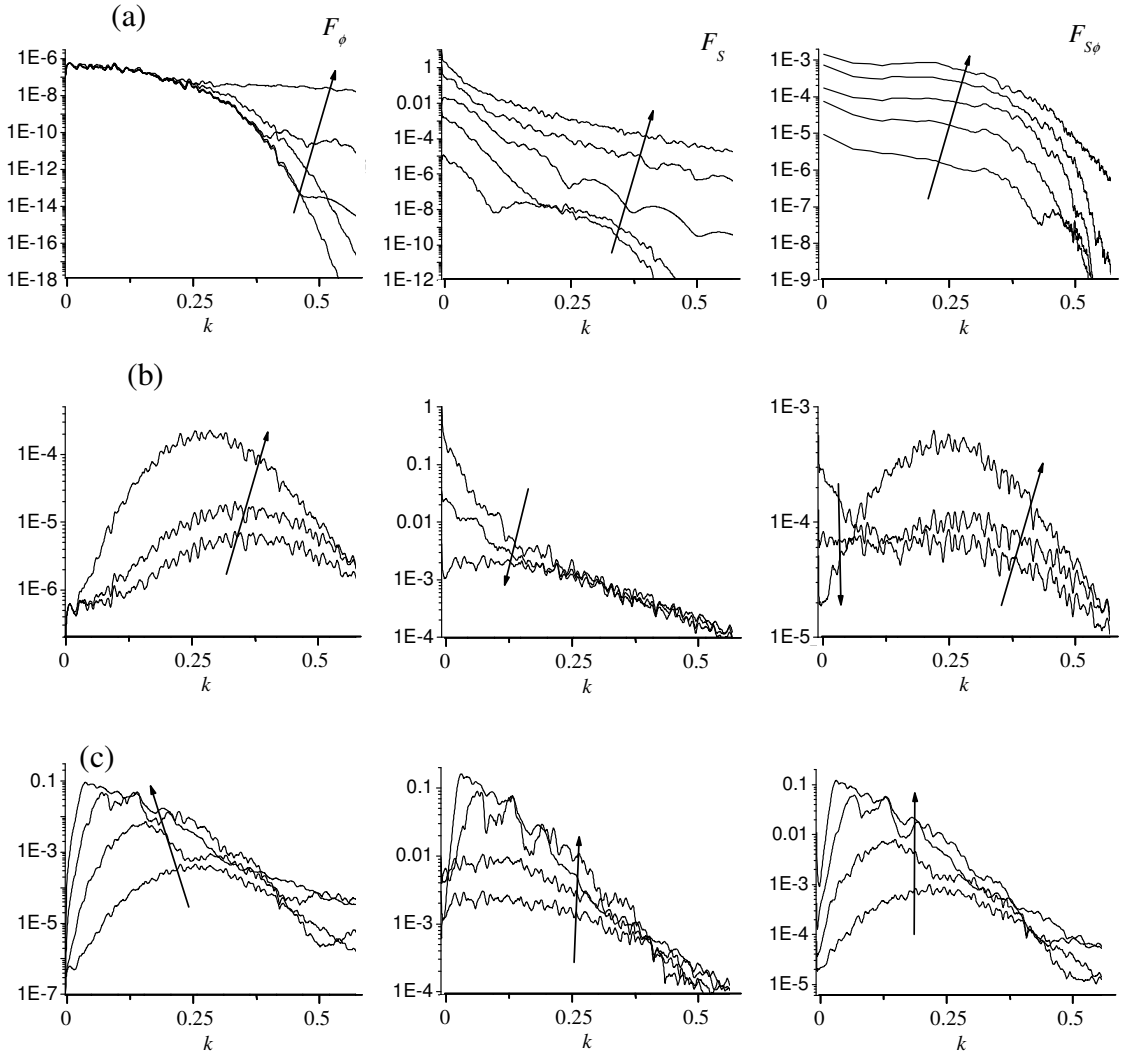

**Figure A1**. Structure factors for MR = $10^{-3}$, corresponding to the following times (increasing in the direction of the arrow): (a) 120, 190, 200, 210, 240; (b) 280,300,450; (c) 550, 3200, 35000, 350000. Each set of results are presented in two figures for clarity.

Figure A2 shows the structure factors for $M_R=1$, corresponding to the profiles shown in figure 7b. As it can be seen in figure 7b, nematic domains are not evenly distributed in space as soon as they form, but some isolated, localized, clearly distinguishable domains of nematic phase formed first. As the first of these domains appears, the concentration profile resembles a pulse function, and consequently the structure factor is a periodic function. As more domains are formed the oscillations in the structure factor smooth out and one maximum becomes distinguishable.

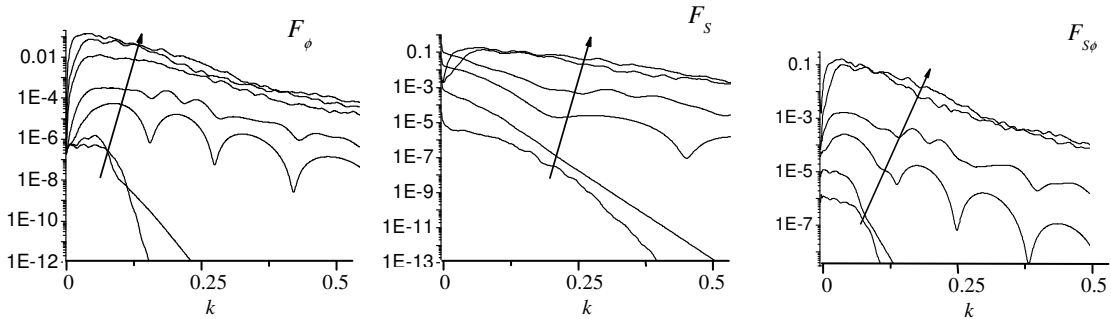

**Figure A2**. Structure factors for MR = 1, corresponding to the following times (increasing in the direction of the arrow): 40, 180, 190, 200, 335, 1750.

Figure A3 shows the structure factors for $M_R=10^3$, corresponding to the profiles shown in figure 7c. For small times the concentration structure factor increases with a maximum located at constant $k_m$. When the nematic structure factor starts to evolve, it is coupled to the concentration structure factor showing the same secondary maxima, but the absolute maximum is located at a higher $k_m$ (figure A3.a), indicating that the structure of fluctuations within the high-concentration domains is dominant. At $t=15$, maximum at smaller $k_m$ becomes the absolute maximum indicating that the domain structure becomes dominant. But as discussed in section 3, this maximum is actually composed by two superimposed peaks, so as time evolves it broadens and moves to higher $k$. Finally for $t = 27$ the two peaks become deconvoluted and the peak located at smaller $k$ becomes dominant.

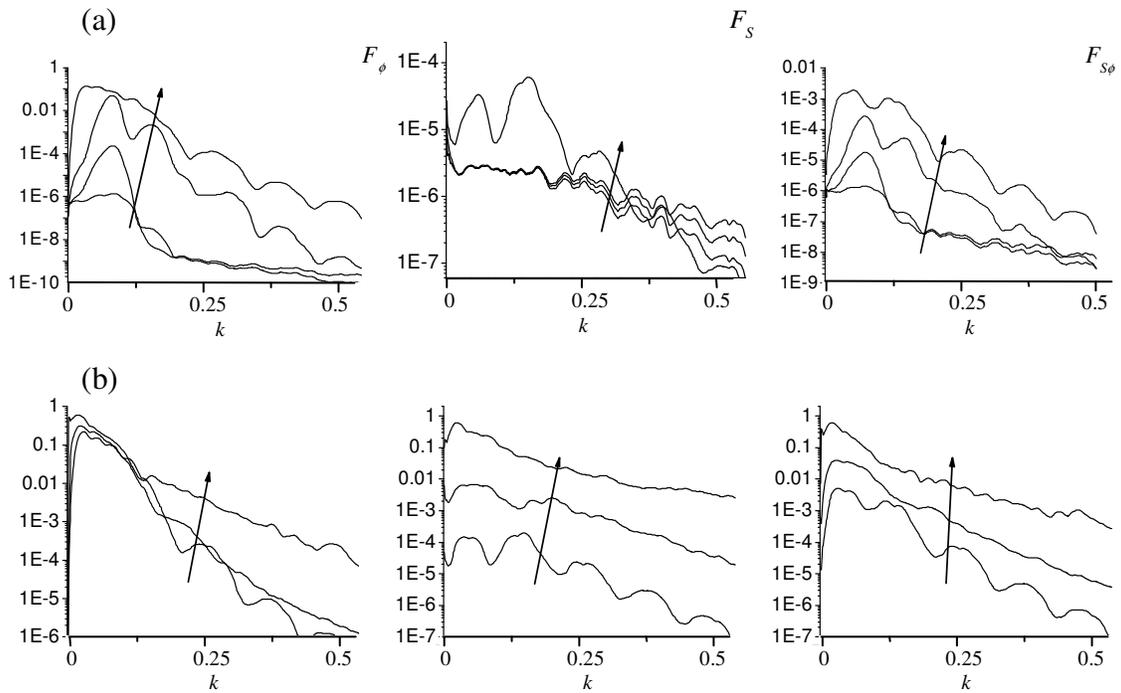

**Figure A3**. Structure factors for MR = $10^3$, corresponding to the following times (increasing in the direction of the arrow): (a) 0.2, 1, 2, 6; (b) 9, 22, 36. Each set of results are presented in two figures for clarity.